\documentclass[twocolumn,prb]{revtex4}

\begin{document}

\title {Coherent control of intersubband optical bistability in quantum wells}
\author {H. O. Wijewardane and C. A. Ullrich}
\affiliation {Department of Physics, University of Missouri-Rolla, Rolla, Missouri 65409}
\date{\today}

\begin{abstract}
We present a study of the nonlinear intersubband (ISB) response of conduction electrons in a 
GaAs/Al$_{0.3}$Ga$_{0.7}$As quantum well to strong THz radiation,
using a density-matrix approach combined with time-dependent density-functional theory.
We demonstrate coherent control of ISB optical bistability, using THz control pulses
to induce picosecond switching between the bistable
states. The switching speed is determined by the ISB relaxation and 
decoherence times, $T_1$ and $T_2$.
\end{abstract}

\maketitle


Intersubband (ISB) transitions in semiconductor quantum wells take place on a meV energy scale and are therefore
attractive for THz device applications. \cite{book} Many ISB effects of practical interest occur in 
the nonlinear regime, such as second- and third-harmonic generation, \cite{heyman1} intensity-dependent
saturation of photoabsorption, \cite{julien,craig} directional control over photocurrents, \cite{dupont}
generation of ultrashort THz pulses, \cite{heyman2} plasma instability, \cite{bakshi}
or optical bistability.\cite{seto,stockman} Inspired by the photoabsorption experiments by Craig {\em et al.}, 
\cite{craig} this Letter presents a theoretical study of
the optical bistability region in a strongly driven, modulation $n$-doped GaAs/Al$_{0.3}$Ga$_{0.7}$As 
quantum well. We here demonstrate, for the first time to our knowledge,
that ISB bistability can be manipulated on a picosecond time scale by short THz control pulses.
This opens up new opportunities for experimental study of optical bistability,
which in turn may lead to new THz applications such as high-speed all-optical modulators and switches.

Most previous theoretical studies of nonlinear ISB dynamics 
were based on density-matrix approaches, either in Hartree approximation \cite{zaluzny,galdrikian,batista}
or using exchange-only semiconductor Bloch equations. \cite{nikonov,castro} 
In this work, we use a density-matrix approach combined with time-dependent density-functional theory,
which has the advantage of formal and computational simplicity, while including exchange-correlation (xc) many-body effects.

We describe the conduction subbands in effective-mass approximation, where $m^* = 0.067m$ and 
$e^* = e/\sqrt{\epsilon}$, $\epsilon=13$, are the effective mass and charge for GaAs. 
Initially, the conduction electrons are in the ground state, characterized by 
single-particle states of the form $\Psi^0_{j{\bf q}_{||}}({\bf r}) =
A^{-1/2} \: e^{i {\bf q}_{||} {\bf r}_{||}} \psi^0_j (z)$, with ${\bf r}_{||}$ and ${\bf q}_{||}$ in the $x-y$ plane.
The envelope function for the $j$th subband $\psi^0_j (z)$ follows self-consistently from a one-dimensional 
Kohn-Sham (KS) equation,\cite{ullrichvignale} with the ground-state density 
\begin{equation}\label{1}
n(z) = 2 \sum_{j,q_{||}}
|\psi_j^0 (z)|^2 \: \theta(\varepsilon_F -E_{jq_{||}}) \:.
\end{equation}
Here, $E_{jq_{||}} = \varepsilon_j + \hbar^2 q_{||}^2/2m^*$, and $\varepsilon_j$ and $\varepsilon_F$ are
the subband  and Fermi energy levels. We choose the electronic sheet density $N_s$ such that
only the lowest subband is occupied, in which case $\varepsilon_F = \pi \hbar^2 N_s/m^* + \varepsilon_1$.

Under the influence of THz driving fields, linearly polarized along $z$, 
the time-dependent states have the form $\Psi_{j{\bf q}_{||}}({\bf r},t) =
A^{-1/2} \: e^{i {\bf q}_{||} {\bf r}_{||}} \psi_j (z,t)$, with initial condition $\Psi_{j{\bf q}_{||}}({\bf r},t_0)
= \Psi^0_{j{\bf q}_{||}}({\bf r})$. In the absence of disorder and phonons, the 
time evolution of the envelope functions follows from a time-dependent KS equation:
\begin{equation} \label{2}
i \hbar\frac{\partial}{\partial t} \psi_j (z,t) = h(t) \psi_j (z,t) \:,
\end{equation}
with 
\begin{equation}\label{3}
h(t) = -  \frac{\hbar^2}{2m^*} \frac{\partial^2}{\partial z^2}
+ v_{\rm dr}(z,t)+ v_{\rm conf}(z)  +  v_{\rm H}(z,t) +v_{\rm xc}(z,t) \:.
\end{equation}
Here, $v_{\rm dr}(z,t) = e{\cal E} z f(t)\sin(\omega t)$ describes the 
driving field, with electric field amplitude $\cal E$,
frequency $\omega$,  and envelope $f(t)$.
$v_{\rm conf}(z)$ is the confining square-well potential, the Hartree potential follows 
from $d^2 v_{\rm H}(z,t)/dz^2 = - 4\pi {e^*}^2 n(z,t)$, and we use the time-dependent
local-density approximation (TDLDA) for the xc potential:
$v_{\rm xc}(z,t) = [d e_{\rm xc}(\tilde{n})/d\tilde{n}]_{\tilde{n}=n(z,t)}$ ($e_{\rm xc}$ is the
homogeneous electron gas xc energy density).
The time-dependent density $n(z,t)$ follows by substituting  $\psi_j(z,t)$ in Eq. (\ref{1}).

To account for disorder or phonon scattering, Eq. (\ref{2}) is replaced
by a density-matrix approach. We expand the first conduction subband as
$\psi_1(z,t) = \sum_{k=1}^{N_b} c_k(t) \psi_k^0(z)$. The associated
$N_b\times N_b$ density matrix $\rho$ has elements $\rho_{kl}(t) = c_k^*(t)c_l(t)$ and initial condition
$\rho_{kl}(t_0) = \delta_{kl}\delta_{1k}$. 
The time evolution of $\rho$ follows from
\begin{equation} \label{4}
i \hbar \frac{\partial \rho(t)}{\partial t} = [h(t),\rho(t)] - R \:,
\end{equation}
with the relaxation matrix $R_{kl} = \hbar[\rho_{kl}(t) - \rho_{kl}(t_0)]/T_{kl}$.
For simplicity,
$T_{kl} = T_1 \delta_{kl} + T_2 (1-\delta_{kl})$, where $T_1$ and $T_2$ are phenomenological relaxation
and decoherence times.

We consider a 40 nm GaAs/$\rm Al_{0.3}Ga_{0.7}As$ square quantum well with $N_s = 6.4\times 10^{10} \: \rm cm^{-2}$,
following  Craig {\em et al.}\cite{craig}
The lowest subband spacing is $\varepsilon_2-\varepsilon_1 = 8.72$ meV, and the ISB plasmon frequency
is found from linear-response theory \cite{ullrichvignale} as $\hbar\omega_{\rm ISB} = 9.91 $ meV.
The system has 9 bound levels, and we take the lowest five to construct the density matrix ($N_b=5$).
We use the ISB scattering times $T_1 = 40$ ps and $T_2 = 3.1$ ps, consistent with recent measurements of
$T_1$ and $T_2$ for similar systems.\cite{heyman3,williams}

To describe ISB photoabsorption, we propagate Eq. (\ref{4}) in the
presence of THz driving fields, switched on at $t_0$ over a 5-cycle linear ramp and then kept at constant intensity
for several hundred ps. From the dipole moment  
$d(t) = N_s \sum_{kl}^{N_b}\rho_{kl}(t) \int dz\: \psi_k^0(z) z \psi_l^0(z)$ we obtain
the photoabsorption cross section (the dissipated power) 
$\sigma(\omega) \sim \omega \int_t^{t+T} \! \cos(\omega t') d(t')\: dt'$, where $T$ is one cycle of the driving field $v_{\rm dr}$.
Following the switching on of the THz field, $\sigma(\omega)$ fluctuates considerably from one cycle to the next,
but eventually approaches a stable value as the transients settle down.

\begin{figure}
\unitlength1cm
\begin{picture}(5.0,7.2)
\put(-6.8,-12.4){\makebox(5.0,6.0){
\includegraphics{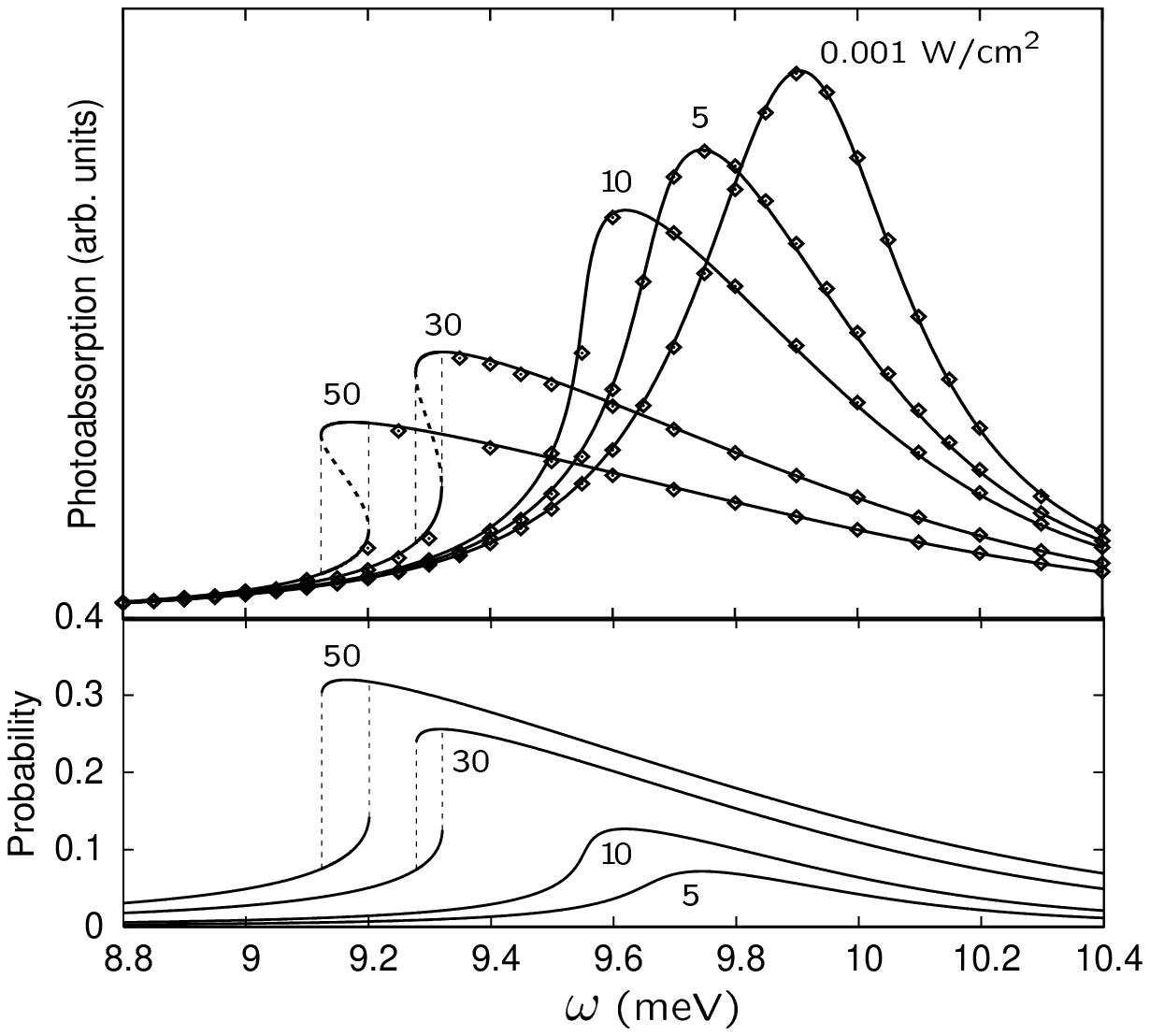}}}
\end{picture}
\caption{\label{figure1} 
ISB photoabsorption $\sigma(\omega)$ and occupation probability $p_2$ of the second subband 
for a 40 nm GaAs/Al$_{0.3}$Ga$_{0.7}$As quantum well with electron density $6.4\times 10^{10} \: \rm cm^{-2}$,
driven by THz fields with intensities as indicated.
Lines: two-level RWA.\cite{zaluzny}
Points: density-matrix calculation [Eq. (\ref{4})].}
\end{figure}

Fig. \ref{figure1} shows $\sigma(\omega)$ calculated with our density-matrix formalism, in good agreement with
a two-level rotating-wave approximation (RWA) model\cite{zaluzny} including xc.
For low intensities, $\sigma(\omega)$ is a Lorentzian with maximum at $\omega_{\rm ISB}$ and width $2\hbar/T_2$. 
Approaching the saturation intensity $I_0 = c\epsilon^{1/2}\hbar^2/8\pi T_1T_2 e^2
|\langle\psi_2^0 | z | \psi_1^0 \rangle|^2$,
the photoabsorption peak shifts to lower energies and changes shape (here, $I_0=26.8\: \rm W/cm^2$).
The physical reason for this effect is that the depolarization shift 
$\Delta = \hbar\omega_{\rm ISB} - (\varepsilon_2-\varepsilon_1)$ is 
proportional to the population difference $p_1-p_2$. Initially, $p_1=1$ and $p_2=0$.
Strong driving fields lead to a decrease of $p_1-p_2$ because of transitions into the 
second and, eventually, higher levels, see the lower panel of Fig. \ref{figure1}. 
Since this population transfer is most efficient around the ISB resonance, the peak of $\sigma(\omega)$ 
shifts more than the tails, leading to an asymmetric line shape.

For intensities greater than 16 $\rm W/cm^2$ we discover regions of optical bistability in $\sigma(\omega)$:
the system has two different ways of responding to one and the same driving field (the middle branch is unstable). 
The two bistable states are characterized by different amplitudes of the density oscillations, 
and different level populations. 

Whether the system will be in the upper or lower bistable state depends on its history. 
A hysteresis-like behavior is observed at fixed intensity and under adiabatically slow frequency changes 
of the driving field:\cite{batista} Entering the bistability region
from the low-frequency side, the system continues on the {\em lower} branch and then 
jumps up at the end of the bistability region. Entering from the other side, the system
follows the {\em upper} branch.
The required continuous, adiabatic frequency tuning of THz driving fields is difficult to realize in experiment,
and of little practical use in exploring ISB optical bistability for potential applications. 
Instead, it would be desirable to switch rapidly between the two bistable states.

In the following, we demonstrate coherent control of ISB optical bistability by short 
THz control pulses. This method is both rapid and robust, and lends itself for experimental implementation.
Fig. \ref{figure1} suggests that a switch from the lower
to the upper bistable state requires a transfer of population into the second subband level.
The necessary energy can be rapidly injected into the system by a short THz pulse. 
On the other hand, switching down from the upper to the lower bistable state requires
upper-level population to relax. We therefore expect this process to be inherently
slower than the up-switch.

\begin{figure}
\unitlength1.0cm
\begin{picture}(5.0,7.)
\put(-4.5,-3.9){\makebox(5.0,6.0){
\includegraphics{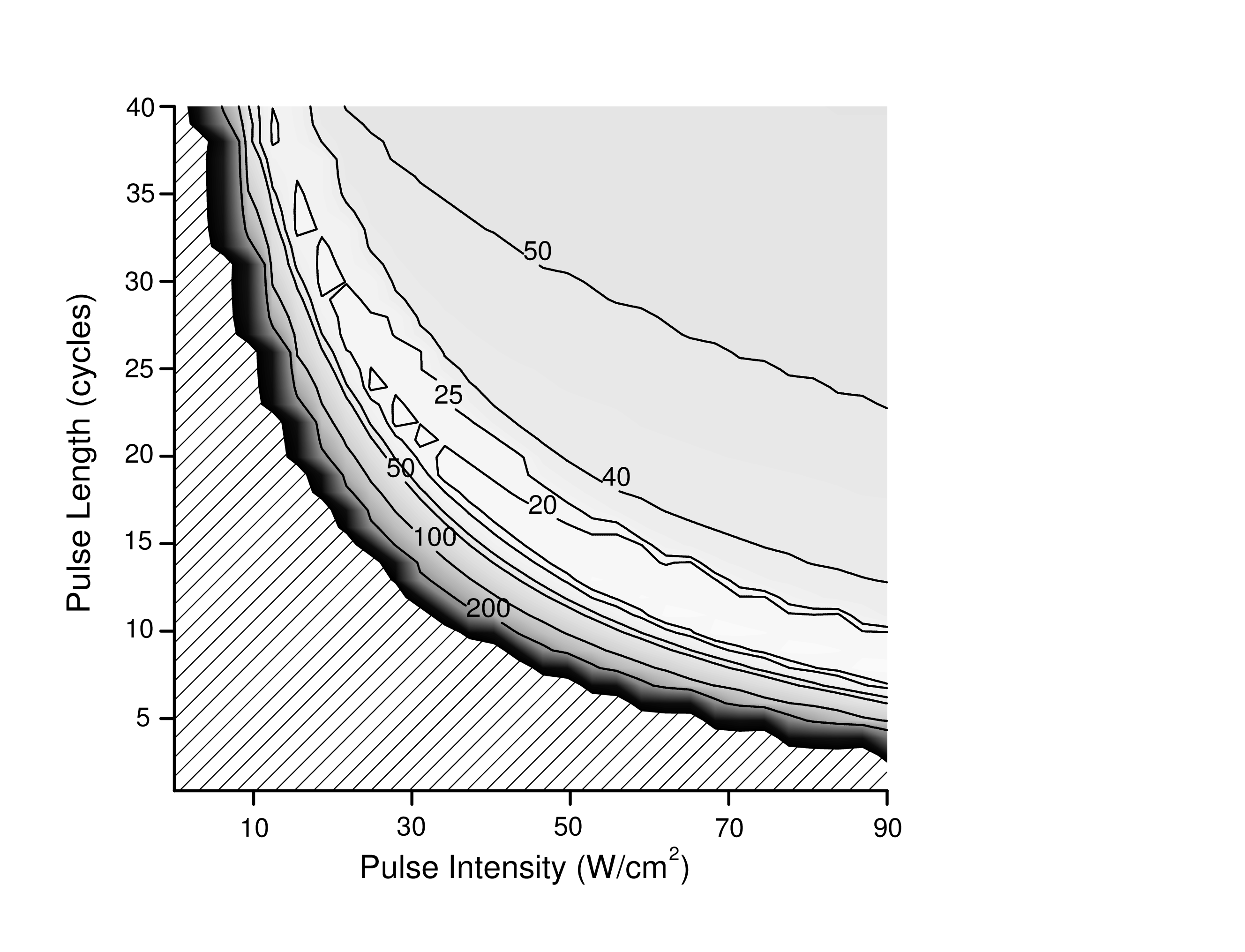}}}
\end{picture}
\caption{\label{figure2} Intensity and length of THz control pulses  used to switch from the lower to the upper
ISB bistable state at $I=30 \:\rm W/cm^2$ and $\omega = 9.3$ meV driving field (0.445 ps per cycle).
Contour labels indicate the switching times in ps. No switching occurs in the hatched region.}
\end{figure}

We consider a driving field with $I=30 \:  \rm W/cm^2$ and 
$\hbar \omega=9.3$ meV (0.445 ps per cycle)
in the middle of the bistable region, such that the
system is in the lower bistable state. At $t_1$, we apply a short pulse 
with the same frequency and {\em in phase} with the driving field, and with a trapezoidal pulse envelope:
linearly turned on and off over 5 cycles,  constant in between over $N_c$ cycles (the precise pulse shape 
is not essential). We calculate $d(t)$ and $\sigma(\omega)$ as above, 
using our density-matrix formalism. After transients and other disturbances induced
by the pulse have subsided, the system either slowly returns to the lower state, or converges towards the
upper bistable state, in which case we define a ``switching time'' as that time after $t_1$ when 
$\sigma(\omega)$ is converged to within 5\% of its final value. 
To study the reverse process, we prepare the system in the upper bistable state and then apply similar
control pulses, {\em phase shifted by $\pi$} with respect to the driving field.
After the transients have settled, the system either remains in the upper state, or goes down to the lower state.

\begin{figure}
\unitlength1.0cm
\begin{picture}(5.0,7.)
\put(-4.5,-3.9){\makebox(5.0,6.0){
\includegraphics{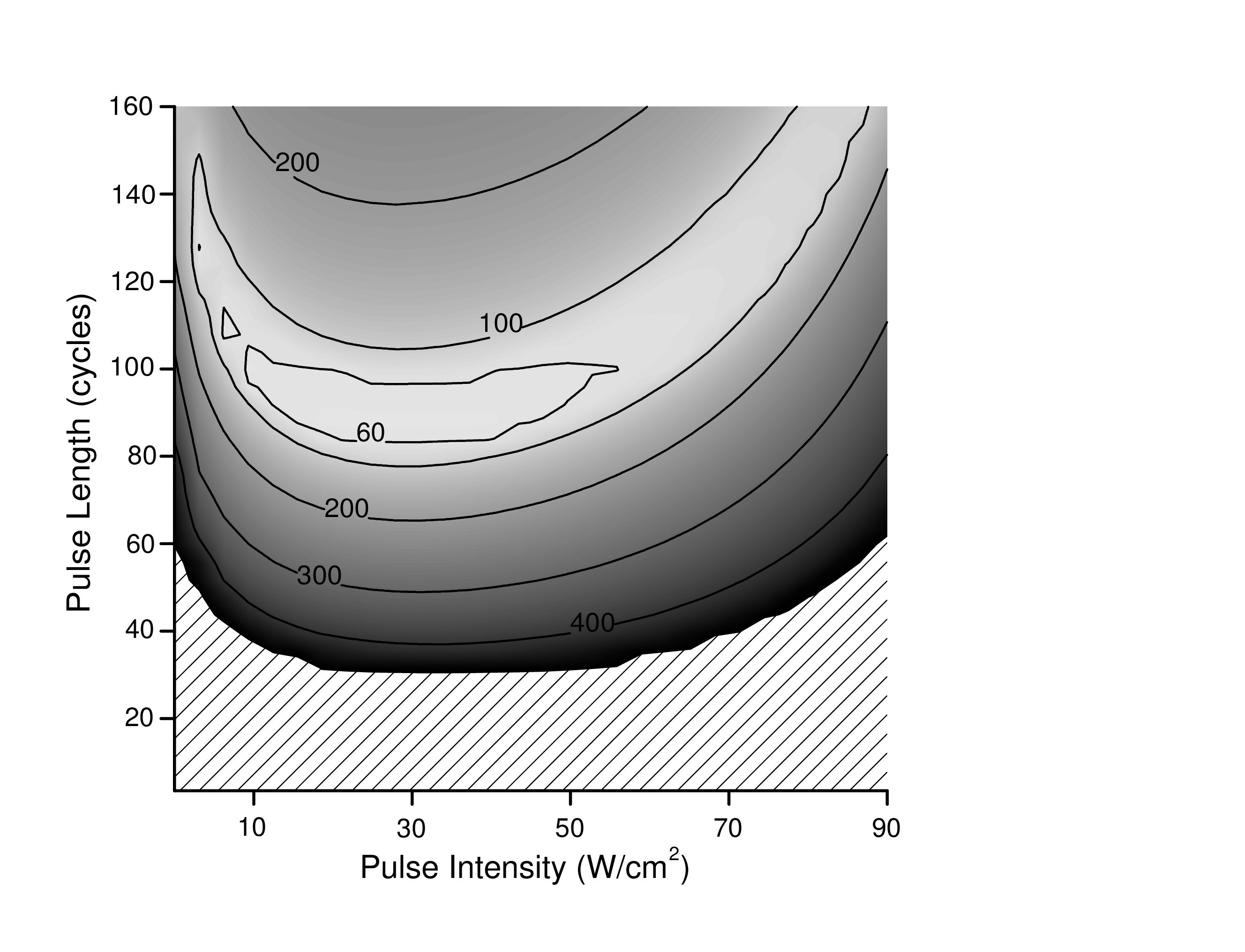}}}
\end{picture}
\caption{\label{figure3} 
Same as Fig. \ref{figure2}, but now for switching from the upper to the lower ISB bistable state.
The THz control pulses are phase shifted by $\pi$ with respect to the driving field.}
\end{figure}

Control pulses of various intensity and length $N_c$ were tested to
determine the conditions for successful up- or down-switching, and to find those pulses that induce
the fastest switching. The results are summarized in Figs. \ref{figure2} and \ref{figure3}.
The hatched areas indicate the unsuccessful pulses: in the case of up-switching,
they do not have sufficient energy to promote enough electrons to the upper level, and in the case of
down-switching, they are too short to allow enough electrons to relax down.

The limiting curve in Fig. \ref{figure2} separating the successful from the unsuccessful pulses 
corresponds to pulses of energy $1.87 \times 10^{-10}\rm \: J/cm^2$. The shortest switching times of 15--20 ps
are achieved for somewhat higher pulse energy, around $3.3 \times 10^{-10}\rm \: J/cm^2$. 
No systematic effort was made to further optimize the control pulses. A physical limit for switching 
speed is set by the decoherence time $T_2$ (here 3.1 ps), which governs the decay of transients.

Down-switch from the upper to the lower ISB bistable state requires longer control pulses,
at least 30 cycles. The associated switching times are limited by the ISB relaxation time $T_1$ (here 40 ps).
We find that the shortest down-switching times of 50--60 ps are achieved by control pulses
with intensity not far from the driving field, $30 \: \rm W/cm^2$. Due to the phase difference
$\pi$, these control pulses effectively suppress the driving field, giving electrons time to relax from the upper level.

In summary, we have simulated coherent control of ISB optical bistability in quantum wells, using
short THz control pulses to switch rapidly between the bistable states.
Switching times are in principle limited only by the intrinsic relaxation and
decoherence times. Interestingly, a shorter relaxation time $T_1$ (i.e., more scattering) 
implies faster switching. More realistic simulations might include microscopic theories for 
$T_1$ and $T_2$ and their intensity dependence, finite temperatures, and a detailed  modeling of the
absorption profile and waveguide geometry of the device. Ultimately, ISB optical bistability
may pave the way towards new THz active optical elements.

We acknowledge support from the donors of the Pe\-tro\-le\-um Research Fund, administered by the ACS,
as well as from the University of Missouri Research Board.


\end{document}